\documentclass[pop,aps,floats,preprint,graphics,amsmath,amssymb,superscriptaddress,showpacs,pre]{revtex4-1}
\usepackage{graphics,graphicx,color}
\begin{document}

\title{Characterisation of the L-mode Scrape Off Layer in MAST: decay lengths}

\author{F. Militello} 
\author{L. Garzotti} 
\author{J. Harrison}
\author{J.T. Omotani}
\author{R. Scannell}
\author{S. Allan}
\author{A. Kirk}
\author{I. Lupelli}
\author{A.J. Thornton and the MAST team}
\affiliation{CCFE, Culham Science Centre, Abingdon, Oxon, OX14 3DB, UK}

\begin{abstract}

This work presents a detailed characterisation of the MAST Scrape Off Layer in L-mode. Scans in line averaged density, plasma current and toroidal magnetic field were performed. A comprehensive and integrated study of the SOL was allowed by the use of a wide range of diagnostics. In agreement with previous results, an increase of the line averaged density induced a broadening of the midplane density profile. This increase was not correlated with divertor detachment, as confirmed by the systematic increase of the target ion flux and decrease of the $D_\gamma/D_\alpha$ emission. Also, no clear correlation is found with the density of the neutral particles at the wall. At comparable density levels, discharges with higher current did not show broadening. Outer target ion saturation current and heat flux decay lengths were measured and compared with midplane data. For the saturation current, the upstream projections of the target values, based on diffusive models, did not match the midplane measurements, neither in amplitude nor in trend, while agreement was found for the heat fluxes, suggesting a different perpendicular transport mechanism for the two channels. Furthermore, the value of the heat flux decay length was quite insensitive to changes in the thermodynamic conditions, in agreement with recent scaling laws. In all the cases studied, sawtooth oscillations were present but they simply rescaled self-similarly the target profiles. The separatrix conditions changed significantly during a sawtooth cycle, but the heat flux decay length and divertor spreading factor remained nearly constant, indicating that these quantities are rather insensitive to the upstream thermodynamic state of the SOL. 

\end{abstract}

\maketitle

\section{Introduction}

As magnetic fusion research progresses towards reactor relevant conditions, it becomes clearer and clearer that the plasma exhaust and its consequent surface interaction will strongly constrain the operational space and determine whether the next generation machines will be successful \cite{Loarte2007}. The level of interaction of the plasma with the solid structures surrounding it is largely determined by the properties of the Scrape-Off Layer (SOL), a narrow region outside the separatrix which is magnetically connected with the divertor target or the walls of the machine \cite{StangbyBOOK}. One of the defining features of the SOL is its width, often described as the decay lengths of some combination of thermodynamic quantities. 

Of particular interest are the particle decay length, $\lambda_n\equiv (\partial \log n/\partial r)^{-1}$, which captures the intensity of the particle fluxes towards the walls and $\lambda_q\equiv (\partial \log q_\parallel/\partial r)^{-1}$, which determines the divertor area over which the power is released (here $n$ is the plasma density, $q_\parallel$ the parallel heat flux and $r$ the radial coordinate). Due to the large difference between parallel and perpendicular transport, the latter being much less efficient, the SOL width is significantly smaller than the machine size, i.e. $\lambda_q\ll R$ and $\lambda_n\ll R$, where $R$ is the major radius. 

Despite the small size of the SOL, the particle flux at the wall, mediated by filamentary structures \cite{Antar2001,Antar2003}, can still be significant. As several experiments have shown over the last 15 years, this is particularly true in high density operations \cite{LaBombard2000,LaBombard2001,Whyte2005,Garcia2007,Carrallero2014} when a shoulder forms in the density profile. Importantly, if ions in the filaments maintain an energy above the sputtering threshold when they reach the solid surfaces they can induce localised erosion \cite{Antar2003}, thus degrading the machine and releasing impurities that pollute the plasma. Also, the density profile in the SOL affects the coupling of the RF antennas, thus determining the efficiency of the heating deposition and it affects the plasma fuelling by influencing the neutral penetration in the core.

Because of the small size of the SOL, the surface interaction area at the divertor is comparably small, thus leading to unacceptable transient and steady state heat loads in high power machines, which entails the need for control techniques (e.g. tailoring of the divertor geometry, impurity seeding, advanced configurations). Similarly, volumetric losses (such as radiation) which would alleviate the divertor loads are limited by the SOL volume which is small compared to the total plasma volume.

In the last few years, an important experimental effort was devoted towards the empirical understanding of the scaling of the heat flux decay length with respect to plasma parameters. This work, based on infra-red thermograpy, resulted in multi-machine regression analyses which suggested that in attached H-mode $\lambda_q$ (at the outer midplane, extrapolated from the divertor data) mainly depends on the inverse of the poloidal magnetic field \cite{Eich2011,Makowski2012,Eich2013}. Interestingly, following L-mode studies \cite{Scarabosio2013} showed remarkably similar trends.   

It is important to note that the SOL width and its features depend on the position, as one moves along the field lines from upstream (i.e. at the midplane) to downstream (i.e. at the target) \cite{Militello2011}. The decay lengths change because of the magnetic configuration (i.e. the flux expansion) and because of the variations in the cross field and parallel transport, which depend on the plasma parameters. 

In this paper, we report a detailed characterisation of the L-mode Scrape Off Layer in the Mega Ampere Spherical Tokamak (MAST). Our study benefits from a wide range of diagnostics, which allowed us to correlate the outer midplane SOL to the divertor region, thus providing useful insight in the parallel effects. In addition, we extended our investigation to the inner divertor region, which behaves in a remarkably different way with respect to its outer counterpart. We devoted particular attention to the investigation of the mechanisms that induce the density profile broadening upstream and how this translates in the downstream region. 

\section{Experimental conditions and diagnostics}

All the results presented here were obtained in MAST \cite{Skyes2001}, which is a small aspect ratio machine with major radius $R\approx 0.9m$ and minor radius $a\approx 0.6m$. Other important peculiarities of the machine are the particularly open divertor and the large distance between the separatrix and the low field side wall ($\sim 0.5 m$). It is also useful to remember that the MAST solenoid fringing field causes the outer strike point to sweep the divertor target by tens of centimetres during the plasma discharge. 

We investigated Ohmic L-mode plasmas in attached conditions over a wide range of upstream SOL conditions. All the discharges were performed with double null diverted configurations. It is important to remark that, due to the particular divertor and wall geometry of MAST, detachment is particularly difficult to achieve \cite{Harrison2011} and none of the results presented here was in that regime. 

\begin{figure}
\includegraphics[height=12cm,width=12cm, angle=0]{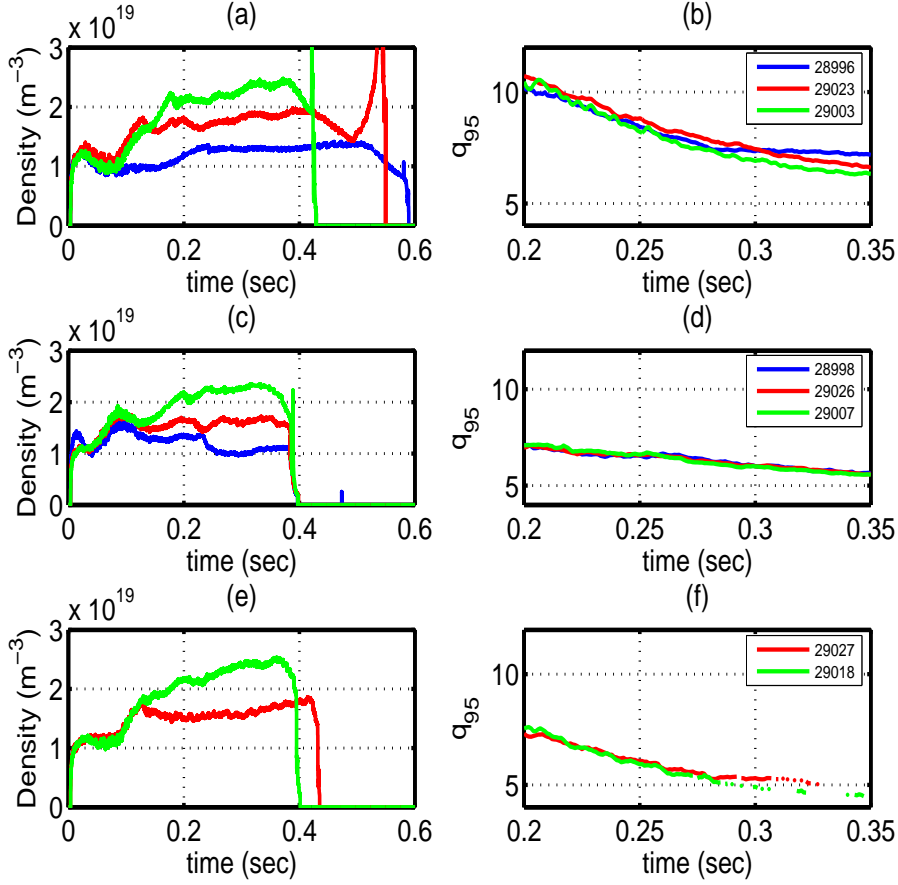}
\caption{Line averaged density and $q_{95}$ for the reference [(a) and (b)], high current [(c) and (d)]and low magnetic field [(e) and (f)]cases.}
\label{fig1}
\end{figure}

Our reference case has $B_T=0.585T$ and $I_p=400kA$, while the line averaged density during the flat top was varied between three levels, $\overline{n}_e\approx [1.3,1.65,2.2]\times 10^{19} m^{-3}$. This was done by varying the deuterium injection from piezo valves situated in the lower and upper divertor regions and at the high field side wall. We also performed a set of density scans at higher current, $I_p=600kA$, with reference field and one at low magnetic field, $B_T=0.4T$, with reference current, for a total of eight different experimental conditions (the low density, low magnetic field case was not suitable for the analysis due to large MHD instabilities). In Fig.\ref{fig1} the line averaged density (measured by the MAST interferometer) and the edge safety factor, $q_{95}$, are shown for the reference, high current and low filed cases.

The divertor region is well monitored with 570 fixed Langmuir probes (FLP) embedded in the centre column (where the inboard strike point lies) and target structures. The probes cover the upper and lower strike points and are $3mm$ apart on the inboard side and $10mm$ on the outboard (the time resolution reaches a maximum of $1kHz$). In addition, the target heat fluxes are measured with two Infra Red cameras (IR) \cite{DeTammerman2010}. The first looks at the upper strike points (inboard and outboard) in the long wave infra red (7.6-8.9$\mu m$) with a spatial and temporal resolution of $6.26 mm$ and $2.5kHz$, while the second looks at the lower strike points (inboard and outboard) in the medium wave infra red (4.5-5$\mu m$) with a spatial and temporal resolution of $5.9mm$ and $0.33kHz$. 

Upstream, electron density and temperature are measured by a high resolution Thomson scattering system (HRTS) \cite{Scannell2010} capable of resolving the core and edge profiles at $0.24kHz$ with $10mm$ radial resolution. Furthermore, the ion saturation current and electrostatic plasma potential were monitored by a mid-plane reciprocating probe with a Mach head mounted on it \cite{Yang2003}. During each discharge, only one probe reciprocation was performed.

In addition, spectrometers measured the $D_\alpha$ ($656.1nm$), $D_\gamma$ ($434.1nm$) and $C_{II}$ ($514nm$) light emission integrated over tangential and radial views at the outer mid-plane and in the divertor regions. The sampling frequency of these measurements was $50kHz$. To complement these measurements, a $D_\alpha$ linear camera, which is a 1024 element 2D CCD array, provided the radial profile of the emission at the midplane. Finally, we used a standard Bayard-Alpert ion gauge, absolutely calibrated and equipped with a fast data acquisition system, to measure with a time resolution of 0.1 ms the pressure of neutral deuterium at the outer midplane. It should be noted that, since the ion gauge is located in a recess shielded from the tokamak magnetic field, its response has a delay of a few milliseconds with respect to the variation of the neutral pressure in the vessel, due to the finite conductivity of the pipe.

\section{Phenomenological description}

In this Section, we discuss the general features of the eight cases analysed, in order to better interpret the quantitative data associated to the SOL decay lengths. All the measurements discussed below were taken in the time window $0.25-0.3sec$ unless otherwise stated. In this period, for each discharge the main plasma parameters had only small variations, e.g. the relative change between highest and lowest value was between $10-20\%$ for $q_{95}$ and between $5-10\%$ for $\overline{n}_e$. The position of the separatrix at the midplane, where the upstream temperatures are evaluated, is estimated using an optically constrained magnetic reconstruction obtained with the EFIT code \cite{Lao1985}, which typically has a 1cm accuracy.      

For convenience, the discharges are clustered in "reference", "high current" and "low field" sets. In the last two subsections we compare the discharges in order to provide a more integrated overview of the plasma conditions, in particular at the divertor and low field side wall.  

\subsection{Reference set}

At the low density level, the Greenwald fraction is $f_{GW}=0.36$. The upstream electron temperature obtained from the HRTS is $T_{e,u} \approx 20eV$, as confirmed by more precise measurements obtained with a retarding field energy analyser in a similar discharge \cite{Elmore2012}, which also suggest a separatrix ion temperature twice as high. The weak thermal coupling between ions and electrons is due to the low collisionality of the upstream SOL. The power crossing the separatrix, $P_{SOL}$, obtained by subtracting the time variation of the energy and the core radiated power measured by the MAST bolometer from the Ohmic power, is between 350kW and 400kW during the flat top. Using a simple two point model to estimate the upstream temperature, we obtain: $T_{e,u} \approx \left[T_{e,t}^{7/2}+\frac{7}{4}\frac{P_{SOL}L_\parallel}{2\pi R \lambda_q (B_p/B_T)\kappa_{0,e}}\right]^{2/7} \approx 20eV$, in agreement with our direct measurements. In this calculation we have used $(B_p/B_T)\approx 0.3$ and $\lambda_q\approx 0.02m$ as discussed in Section \ref{outer_target} and $\kappa_{0,e}$ is the temperature independent part of the parallel electron conductivity. Also, this temperature agrees very well with the empirical scaling $T_{e,u} = 1.4\times 10^{36} \overline{n}^{-1.8}I_p^{1.15}$ (see Eq.4.42 of \cite{StangbyBOOK}), and so do almost all the other seperatrix temperatures given in this and the following Subsections. Since the HRTS data show a separatrix density $n_{e,u}\approx 0.35\times 10^{19} m^{-3}$, and the connection length, $L_{\parallel}$, is around 13m, the electron collisionality can be estimated as $\nu_{*,e}=10^{-16}n_{e,u}L_\parallel/T_{e,u}^2\approx 11$, which would imply a sheath limited regime, but close to the transition to the conduction limited regime. In the outer upper divertor, the FLP measured a peak density $n_{e,t}\approx 0.35\times 10^{19} m^{-3}$ and an electron temperature at the strike point position $T_{e,t}\approx 14eV$. Based on the results of \cite{Asakura1997,Elmore2012}, we expect an equilibration between target electron and ion temperature, which should be roughly the same, i.e. $T_{i,t}\approx T_{e,t}$. Comparing upstream and downstream data, we find pressure conservation along the field lines, which can be formulated as: $n_{e,u}(T_{e,u}+T_{i,u})\approx 2n_{e,t }(T_{e,t}+T_{i,t})$ (see e.g. \cite{Asakura1997}). 

The second density level corresponds to $f_{GW}=0.48$. The higher line averaged density corresponds to an increase in the separatrix value which is $n_{e,u}\approx 0.49\times 10^{19} m^{-3}$ and also leads to a cooling of the upstream temperature, which is around $T_{e,u}\approx 14eV$. The conditions at the upper outer target are $n_{e,t}\approx 0.7\times 10^{19} m^{-3}$ and $T_{e,t}\approx 7eV$, thus showing again pressure conservation in the assumption of similar ion/electron temperature ratios as in the previous case. The upstream electron collisionality is $\nu_{*,e}\approx 32$, which would suggest a conduction limited regime, compatible with the fact that the target temperature is half the upstream. 

The third level, at $f_{GW}=0.6$, shows a similar upstream temperature, $T_{e,u}\approx 10eV$, as the intermediate density case, which is reasonable considering a conduction limited regime and comparable power crossing the separatrix, while the separatrix density reaches $n_{e,u}\approx 0.55\times 10^{19} m^{-3}$. This is again compatible with the measurements in \cite{Elmore2012}, which also indicate an upstream ion temperature a factor two higher. At the target, $n_{e,t}\approx 0.75\times 10^{19} m^{-3}$ is measured, while the temperature data show a large scatter. Based on the results in \cite{Elmore2012}, we estimate the electron and ion target temperature for this case at around 5eV. Like in the previous cases, pressure is conserved and the upstream collisionality for this density level is $\nu_{*,e}\approx 71$.   

\subsection{High current set}

\begin{figure}
\includegraphics[height=12cm,width=12cm, angle=-90]{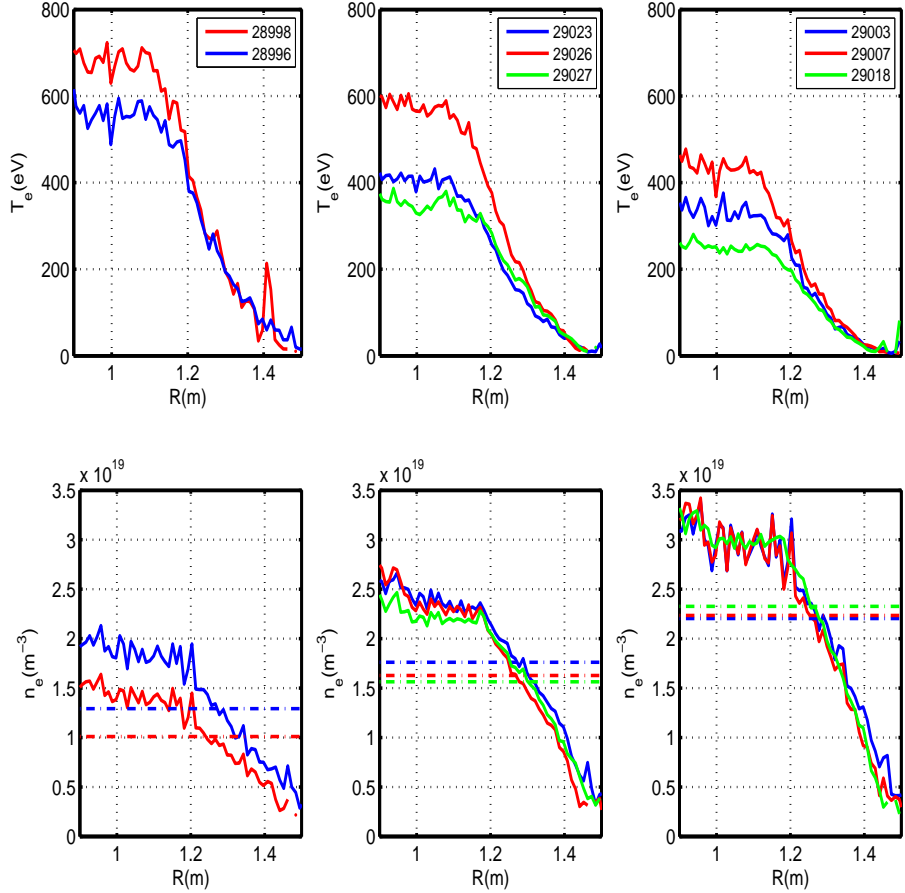}
\caption{Time averaged profiles of the electron temperature (upper row) and density (lower row) for the 8 discharges during the flat top. Only the outboard half is shown. From left to right, the columns show increasing density levels. Blue, red and green curves represent the reference, high current and low field sets. In the lower row, the horizontal dash-dot lines represent the values of the line averaged density of the different discharges.}
\label{fig2}
\end{figure}

The second set of discharges was performed at a higher current ($I_p\approx 600kA$) and at the reference magnetic field. The main consequences of this change are a larger Ohmic heating, $P_{Ohm}\sim I_p$, which leads to a temperature increase with respect to similar density levels in the reference cases and a reduction of the edge safety factor, $q_{95}\sim I_p^{-1}$ (this also shortens the connection length to roughly 85\% of the reference value). Figures \ref{fig1} and \ref{fig2} show that these effects are observed while the core density profiles are reasonably aligned to each other apart from the low density level, in which the line averaged density is around 20$\%$ lower than the reference discharge.

In the low density level, with $f_{GW}\approx 0.19$, the measured electron density and temperature at the separatrix are $n_{e,u}\approx 0.28\times 10^{19}$ and $T_{e,u}\approx 22eV$. Due to the low collsionality, $\nu_{*,e}\approx 6$, the divertor is expected to be in a sheath limited regime, which is confirmed by the fact that target electron temperature is similar to the upstream value $T_{e,t}\approx 20eV$. For this case we do not have ion temperature measurements but we expect low thermal coupling between ions and electrons, due to the low collisonality. Given the measured electron density at the target, $n_{e,t}\approx 0.3\times 10^{19} m^{-3}$, pressure conservation, which should be a consequence of the sheath limited regime, would be satisfied with $T_{i,u}\approx 3 T_{e,u}$ if $T_{i,t}\approx T_{e,t}$, which is comparable to the values of the reference case. 

The second density level, at $f_{GW}\approx 0.32$, is characterised by $n_{e,u}\approx 0.51 \times 10^{19}$ and $T_{e,u}\approx 21 eV$, which leads to an upstream collisionality $\nu_{*,e}\approx 9$. The target data give $n_{e,t}\approx 0.7 \times 10^{19}$ and $T_{e,t}\approx 15eV$. Pressure conservation would again be satisfied if the upstream ion temperature was roughly three times higher than the electron temperature under the condition that the target thermal coupling was strong. 

Finally, the last density level corresponds to $f_{GW}\approx 0.41$. Upstream and downstream measurements give $n_{e,u}\approx 0.6 \times 10^{19}$, $T_{e,u}\approx 12eV$, $n_{e,t}\approx 1.3 \times 10^{19}$ and $T_{e,t}\approx 7eV$. With these conditions, the divertor should still be in attached conditions and conduction limited, since $\nu_{*,e}\approx 45$. Pressure conservation would require an upstream ion temperature around twice as high as the electron, compatible with higher collisionality with respect to the lower density levels and hence higher thermal coupling.

\subsection{Low field set}

The final set of discharges, which does not include a low density case, was performed at a lower magnetic field, $B_T=0.4T$, and at the reference current. This gave a $q_{95}$ similar to the high current cases since $q_{95}\sim B_T/I_p$ and an Ohmic heating similar to the reference discharges. The core density profiles match reasonably well the reference and high current sets, while the temperature profiles are comparable to the reference cases, albeit lower in the innermost part of the plasma, see Fig.\ref{fig2}.     

As expected, the Greenwald fraction of the intermediate density discharge, $f_{GW}= 0.44$, is similar to its reference case counterpart, and so are the  separatrix density, $n_{e,u}\approx 0.5 \times 10^{19}$, and temperature, $T_{e,u}\approx 12eV$. The associated collisionality, which takes into account the reduction of the connection length due to the lower $q_{95}$, is  $\nu_{*,e}\approx 24$. At the target, $n_{e,t}\approx 0.6 \times 10^{19}$ and $T_{e,t}\approx 9eV$. 

Also in the high density case the matching with the reference case is good as upstream we measure $f_{GW}=0.63$, $n_{e,u}\approx 0.7 \times 10^{19}$ and $T_{e,u}\approx 10eV$, which gives $\nu_{*,e}\approx 49$. Downstream, the Langmuir probes give $n_{e,t}\approx 0.7 \times 10^{19}$ and $T_{e,t}\approx 7eV$. 

\subsection{Radiation, total ion flux and divertor regime}

Using the combined data of the discharges, we can shed some light on their divertor regime and in particular on whether detachment occurred. The first indication of attached conditions comes from the fact that pressure balance seems to be satisfied in all the discharges (see Subsections above), which indicates that only a small amount of momentum is transferred to neutral particles. This estimate is, however, very rough as measurements are sometimes incomplete and subject to significant error bars. While pressure conservation should be carefully checked in the future, we present in the following more solid arguments which suggest absence of detachment.   

\begin{figure}
\includegraphics[height=12cm,width=12cm, angle=-90]{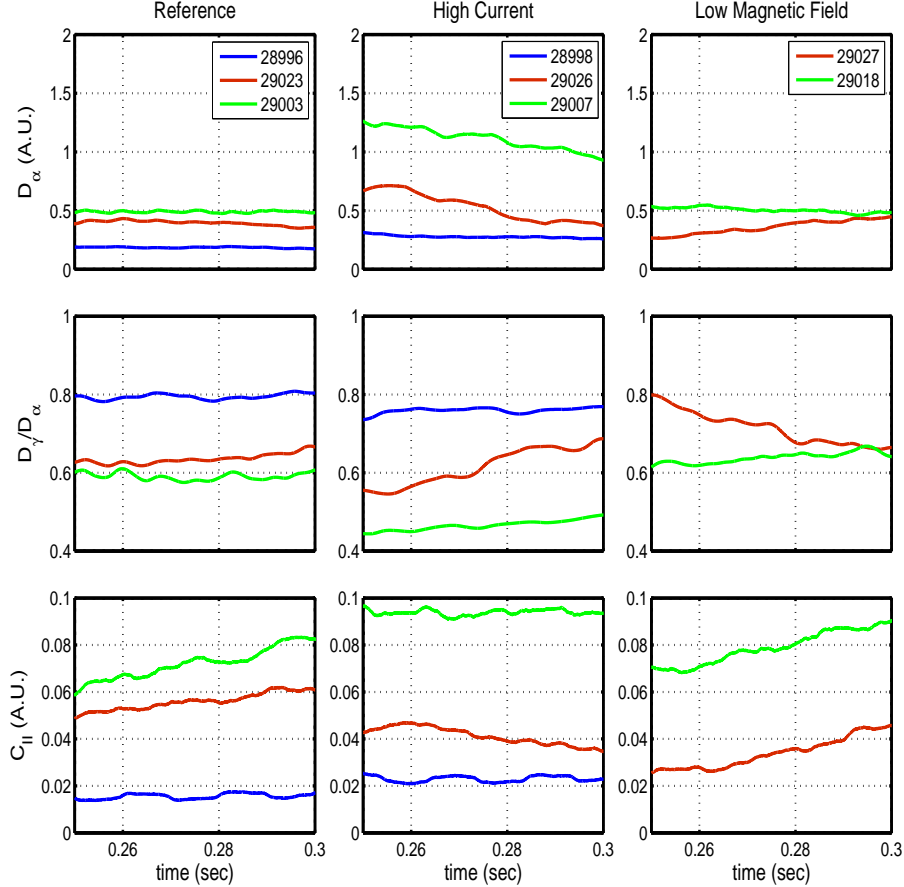}
\caption{Radiation measurements for the 8 discharges. The first row shows the $D_\alpha$ emission, the second the $D_\gamma/D_\alpha$ ratio, the third the $C_{II}$ emission. From left to right, the columns show the reference, high current and low magnetic field discharges. The signals were smoothed with a box-car average with a 10ms window to filter out the sawtooth oscillations. All the measurements were taken from radial views at the upper divertor.}
\label{fig4}
\end{figure}

For each set of discharges (reference, high current and low field), during the flat top phase, a higher line averaged density corresponds to higher core and edge radiation, measured by the bolometer (not shown) and the deuterium and carbon line emission, see Fig.\ref{fig4}. At the same time, the $D_\gamma/D_\alpha$ ratio at the target steadily decreases, as expected when the divertor density increases in conditions in which excitation dominates over recombination, i.e. \textit{before} detachment \cite{McCracken1998}. This can be clearly seen in the plots on the second row of Fig.\ref{fig4} when different density levels of the same set of discharges are compared. It is worthwhile remarking that a reversal of this trend in the reference set was transiently observed in a very high density shot, corresponding to an average $\overline{n}_e\approx 2.9\times 10^{19} m^{-3}$. This discharge could not be included in the present analysis as it showed large oscillations in the plasma parameters and no steady state period due to the difficulty of feedback controlling the density. 

\begin{figure}
\includegraphics[height=12cm,width=12cm, angle=0]{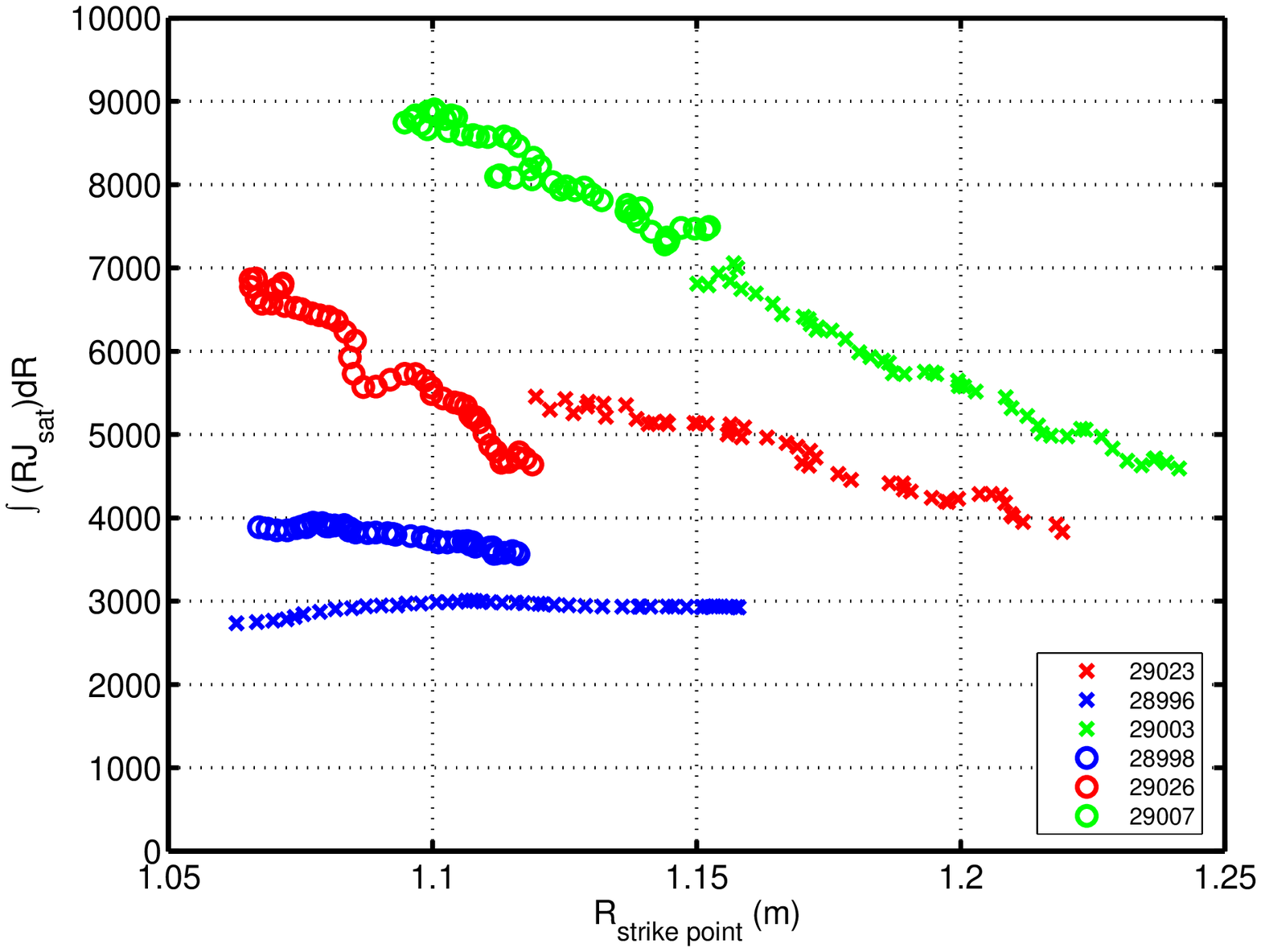}
\caption{Integrated ion saturated current measured by the Langmuir probes, used as a proxy for the total particle flux at the outer target as a function of the strike point position during the flat top phase. Different symbols represent different discharges.}
\label{fig3}
\end{figure}

Coming back to the analysed discharges, the total ion flux to the target, proportional to the radial integral of $n_{e,t}T_{e,t}^{1/2}\sim J_{sat,t}$, was estimated using the divertor Langmuir probes. Larger ion fluxes were observed when the line averaged density was higher, again confirming that the discharges were not detached. This is shown in Fig.\ref{fig3}, where the total ion flux, $\Gamma_i \sim \int J_{sat,t}RdR$ is plotted as a function of the upper outer strike point position during the analysed time window. The reduction of the total ion flux as $R_{Strike Point}$ increases is attributed to the lower plasma temperature at the target as the divertor leg moves outward.

\subsection{Neutral density and plasma sources} \label{neutrals}

\begin{figure}
\includegraphics[height=12cm,width=12cm, angle=0]{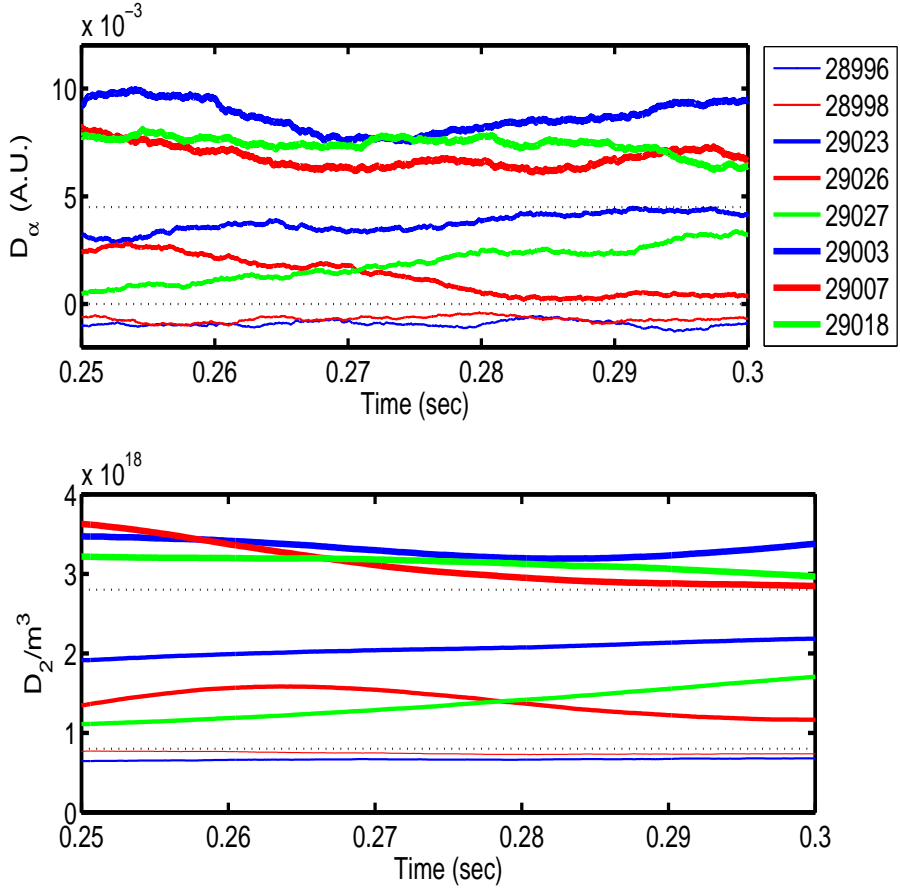}
\caption{Upper panel: time evolution of the midplane radial $D_\alpha$ signal smoothed with a moving average technique (using a $10msec$ window). Lower panel: time evolution of the neutral density at the wall. The horizontal dotted lines show the separation between the three density levels. The oscillations in the signals are attributed to the response of the plasma density control system.}
\label{fig5}
\end{figure}

The neutral pressure at the wall, measured near the midplane by the fast ion gauge, allows us to estimate the mean free path of the neutrals and their likelihood of being ionized before reaching the separatrix. In the discharges presented here, the neutral pressure showed a good correlation with the line averaged density and appeared to be quite insensitive to other operational conditions, e.g. the plasma current or the toroidal magnetic field, see the lower panel of Fig.\ref{fig5}. 

More details on the plasma sources and the neutral density can be obtained by comparing the radially integrated $D_\alpha$ emission at the midplane and the neutral density at the wall. In all the cases treated, the $D_\alpha$ emission is localized around the outer separatrix, as the linear camera measurements show. The radially integrated emission measurements (see Fig.\ref{fig5}) are therefore representative of that region of the plasma. As Fig.\ref{fig5} shows, for both separatrix and wall signal, the strongest dependence is on the plasma line averaged density, suggesting that the particle sources are similar between sets at comparable $\overline{n}_e$. 

It is also useful to mention that previous work \cite{Huang2010} has shown that, for a typical discharge duration, the MAST wall is expected to retain almost all the particles impinging into it.

\section{Upstream profiles and decay lengths} \label{upstream}

\subsection{Ion saturation current}

\begin{figure}
\includegraphics[height=12cm,width=12cm, angle=0]{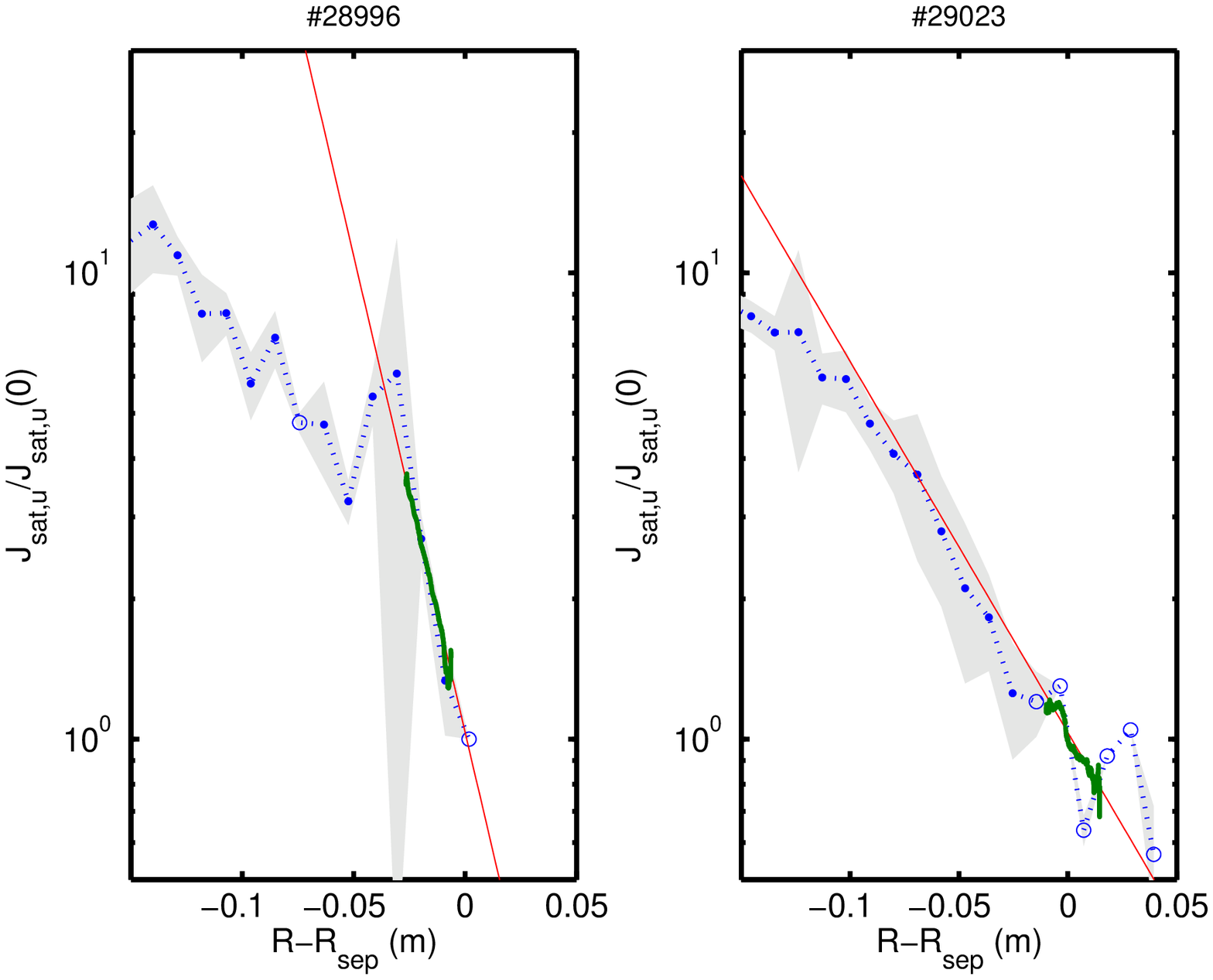}
\caption{Comparison between the normalized ion saturation current HRTS averaged profiles (dashed lines) and RP profiles (thick line). A thin line represents the exponential fit to the RP data used to calculate $\lambda_{J,u}$. The points indicate the position and value of the HRTS channels, with an empty symbol representing less reliable channels that had more than $50\%$ unusable data in the period of the average. The gray area (plotted only for the reliable channels) represents the uncertainty in the HRTS measurements.}
\label{fig6}
\end{figure}

The plasma density and the ion saturation current are strongly related. A Mach probe mounted on a reciprocating arm was used to evaluate the upstream profiles of the latter. High Resolution Thomson Scattering data were in good agreement with the RP, as shown in Fig.\ref{fig6} for the reference discharges, with the ion saturation current estimated as $J_{sat,u}\sim n_{e,u}T_{e,u}^{1/2}$. 

We remark that in the evaluation of the profiles, we processed only data during the first part of the reciprocation, i.e. before the probe started to move \textit{out} of the plasma. This was necessary as the probe reciprocation perturbs the local properties of the plasma so that on its way out, the probe measures steeper profiles. Importantly, if both inward and outward data were averaged and processed together, the flattening in the far SOL would disappear. In practical terms, our data were collected between $0.23-0.28sec$ but with different starting points and window length depending on the discharge (all the time windows had a minimum length of $30msec$). 

\begin{figure}
\includegraphics[height=10cm,width=14cm, angle=0]{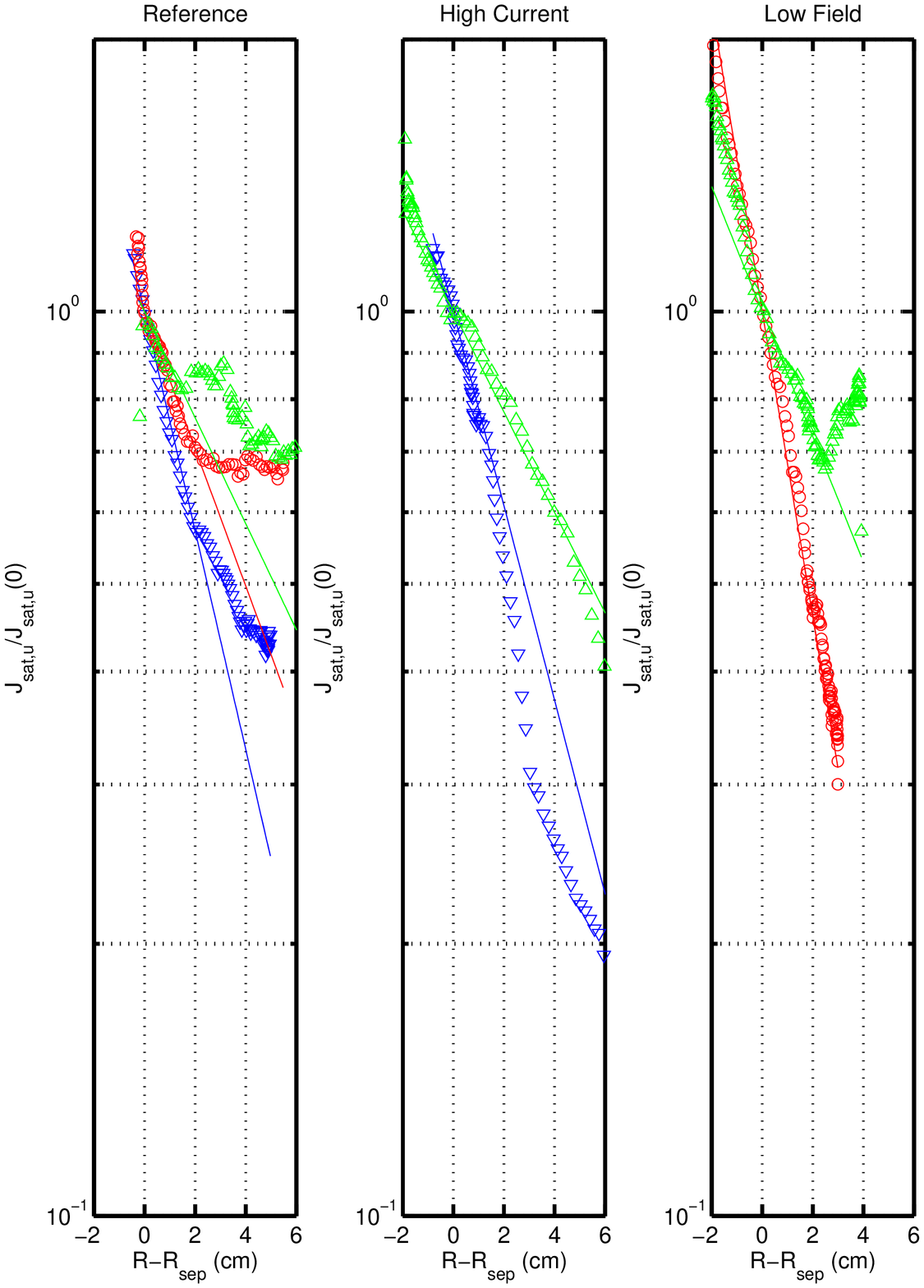}
\caption{Profiles of the ion saturation current normalized to its separatrix value as a function of the distance from the separatrix. The profiles from the RP measurements are smoothed using a moving average with a bin of 15msec. For each set, the symbols ($\color{blue} \triangledown$), ($\color{red} \circ$) and ($\color{green} \vartriangle$) represent the low, intermediate and high density levels respectively. Straight lines, representing the exponential profiles with decay lengths given in Table \ref{tab1}, are plotted for comparison.}
\label{fig7}
\end{figure}

The comparison between the different discharges shows that the SOL profile broadening is consistently correlated with the increase of the line averaged density, as already observed in other machines \cite{LaBombard2001,Lipschultz2002,Garcia2007,Carrallero2014}, see Fig.\ref{fig7}. However, there is a qualitative and quantitative difference between the reference set and the other discharges. In the former, the slope of the profiles is steeper close to the separatrix, while it flattens in the far SOL, especially for large $\overline{n}_e$. Also the high density low magnetic field case showed a similar trend. There is extensive experimental literature documenting this two region feature of the diverted SOL \cite{McCormick1992,Asakura1997,LaBombard2001,LaBombard2005,Lipschultz2005}, although the mechanisms behind this phenomenon remain elusive. In the high current cases, however, the profiles did not show any visible flattening in the far SOL. This is in agreement with the results reported in \cite{McCormick1992} for a closed divertor, which suggests that the current effect is insensitive from divertor geometry and closure. In addition, in most of the cases analysed, the gradients of the profiles in the near SOL and just inside the separatrix did not show sharp discontinuities.  

\begin{figure}
\includegraphics[height=10cm,width=14cm, angle=0]{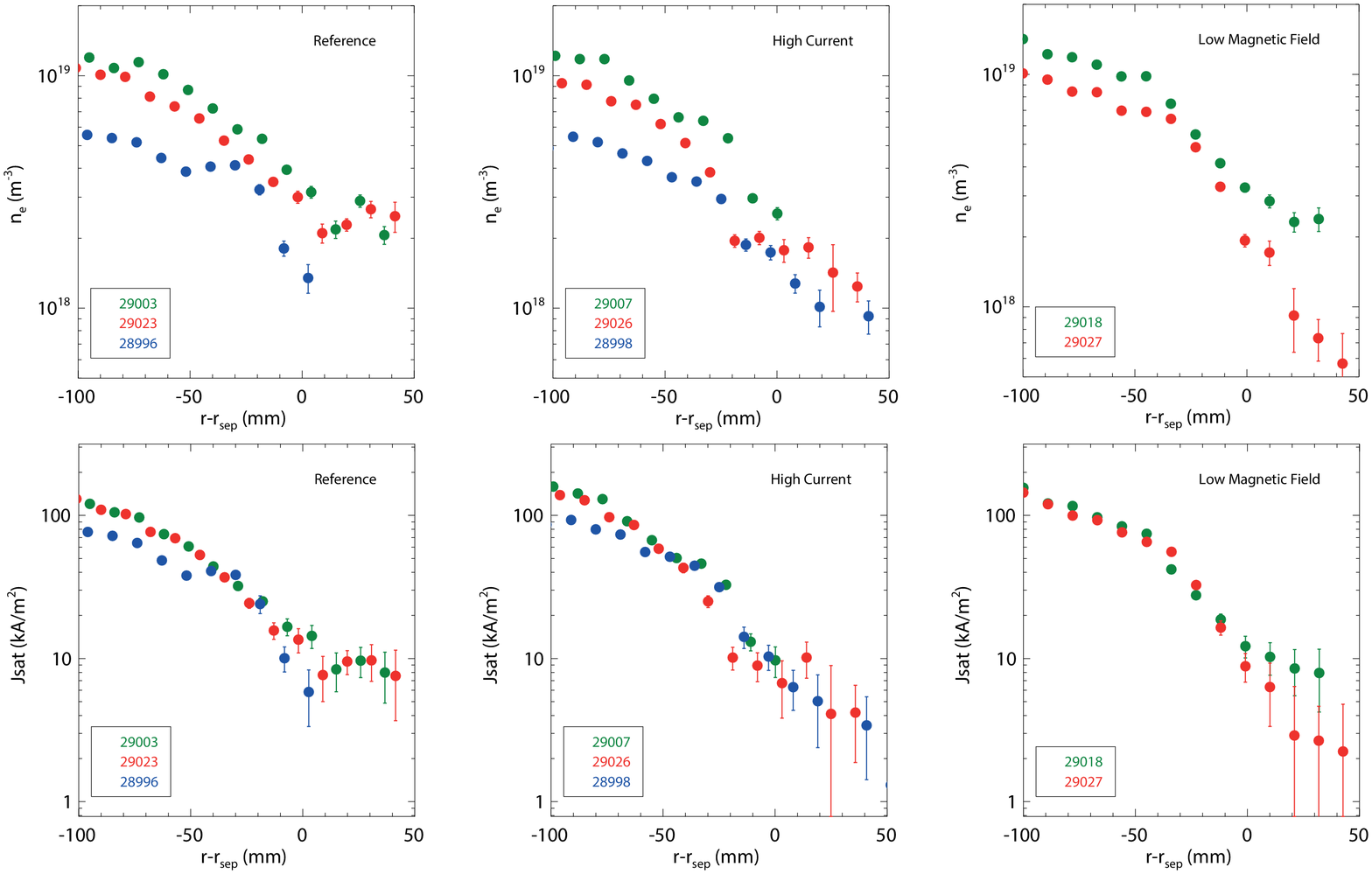}
\caption{Profiles of the density and ion saturation current normalized to their separatrix value as a function of the distance from the separatrix. The profiles are obtained from HRTS measurements with the technique described in the text.}
\label{fig7a}
\end{figure}

These reciprocating probe results were confirmed by measurements taken with the HRTS system, which also reiterated that the density follows the ion saturation current profile \ref{fig7a}. The data shown were obtained by accumulating the scattered signal time traces in a stationary plasma over a period of 50ms. Since the lasers fire at 240Hz, this corresponds to 12 scattered pulses. Following the accumulation process, the signal time traces are fitted to obtain the number photons in each spectral bin of a polychromator, which are in turn fitted to obtain density and electron temperature. While the diagnostic typically measures well down to $2-5\times 10^{18}m^{-3}$ in single pulse mode, accumulating over 12 laser pulses allows us to measure with confidence down to $\sim 0.6-1.4\times 10^{18}m^{-3}$. It is important to accumulate the time traces instead of fitting each scattered signal individually and averaging the integrals, as the signals for an individual time trace are close to the noise floor. It should be noted that our diagnostic has a minimum measurable temperature of approximately 5eV due to the proximity of the final spectral bin to the laser wavelength.

From these measurements, we observe that the decay length in the near SOL ranges between $\sim 3.5cm$ and $\sim 7cm$ in all the experimental conditions examined. The values of $\lambda_J$ with their fit errors are given in Table \ref{tab1}.
\begin{table}
\caption{Ion saturation current decay lengths calculated upstream in the near SOL (in cm).}
\begin{tabular}{|l||c|c|c|}
\hline 
& low $\overline{n}_e$ & intermediate  $\overline{n}_e$ & high $\overline{n}_e$ \\
\hline\hline
Reference &  3.58 $\pm$ 0.01 & 5.72 $\pm$ 0.01 & 7.41 $\pm$ 0.04  \\ \hline
High Current  & 4.05 $\pm$ 0.01 & / & 7.82 $\pm$ 0.02 \\ \hline
Low field  & / & 2.58 $\pm$ 0.03 & 6.27 $\pm$ 0.01 \\ \hline
\end{tabular}
\label{tab1}
\end{table}  
The absence from the table  and from Fig.\ref{fig7} of the intermediate density in the high current set is due to the fact that the reciprocation was not deep enough to evaluate the near SOL features. 

From these results, the dominant parameter that controls the near SOL ion saturation current decay length seems to be the line averaged density. The comparison between the reference and the high current sets also suggests low sensitivity to $I_p$, and hence that the Greenwald fraction is not a good scaling quantity for $\lambda_J$. 

In the far SOL, $\lambda_J$ responds to variations of $\overline{n}_e$ more rapidly than in the near SOL for the reference discharges. Also, differently from the near SOL, plasma current affects the gradients in this region. Previous work \cite{LaBombard2000,Lipschultz2005} suggested that an increase of the ionization sources due to the neutral recycling at the wall could provide a possible explanation for the far SOL broadening. However, the results of Sec.\ref{neutrals} suggest that the level of neutrals depends exclusively on the line averaged density, while the broadening has a clear dependence on $I_p$ and $B_t$ as well. This implies that wall recycling cannot explain the flatter profiles in the far SOL of the reference set.  

In our experiment, we do not observe a connection between broadening and detachment, which is sometimes believed to affect the SOL profiles. While the divertor plasma is changing between different discharges, so is the SOL above the X-point, thus making it extremely difficult to pinpoint the mechanism for the profile flattening to a particular SOL location, which might be due to global properties of the edge plasma. 

\subsection{Parallel heat flux}

In our experiments, no direct measurements of the heat flux profile are available upstream. However, the decay length close to the separatrix can be estimated by extrapolating to the SOL region reasonable combinations of the HRTS and RP data. This is justified by the continuous nature of the gradients across the separatrix observed in the ion saturation current. As a standard approximation \cite{StangbyBOOK}, if conduction is the dominant parallel heat loss mechanism, the heat flux decay length is estimated as a fraction of the electron temperature decay length, $\lambda_q \approx 2/7\lambda_T$. On the other hand, if parallel convection is assumed to be more important, a better approximation is $\lambda_q^{-1}\approx \lambda_n^{-1}+3/2\lambda_T^{-1}$ (from $q_\parallel \sim nT^{3/2}$). Both regimes constitute a simplification of the actual parallel heat loss physics, which at the very least is a combination of both effects. In principle, in perfectly symmetric double null configurations, the parallel heat flux at the midplane should be exactly zero and therefore $\lambda_q$ is an ill defined quantity. Therefore, our measurements are simply reflecting combinations of the temperature and density profiles and as such they should be interpreted. However, it makes sense to use the convective approximation to compare the upstream with the downstream data, since the sheath at the target enforces $q_{\parallel,t}\sim n_{e,t}T_{e,t}^{3/2}$ \cite{StangbyBOOK}. In other words, the convective estimate for our upstream measurements is representative of the one direction heat flux that would be obtained if a solid surface was present at the midplane.

For the sake of completeness, we give here estimates of $\lambda_q$ in both limits, knowing that these values are only indicative, although probably in the right order of magnitude, and that $\lambda_q$ convective is probably more relevant. To improve the accuracy of the estimate, instead of using $\lambda_n$ in the calculation of the convective decay length, we use $\lambda_J$, so that $\lambda_q^{-1}\approx \lambda_J^{-1}+\lambda_T^{-1}$. The upstream decay lengths calculated with this method are summarized in Tab.\ref{tab2}. The errors in the table are a result of the fitting procedure (based on the last 5 channels of the HRTS) and were calculated assuming a 95\% confidence interval. Note also that the temperature decay lengths from the HRTS can be easily extrapolated from the conductive $\lambda_q$.
\begin{table}
\caption{Parallel heat flux decay lengths calculated in the near SOL (in cm). Values extrapolated from HRTS data.}
\begin{tabular}{|l||c|c|c|}
\hline 
& low $\overline{n}_e$ & intermediate  $\overline{n}_e$ & high $\overline{n}_e$ \\
\hline\hline
Reference (cond.)&  0.62 $\pm$ 0.12 & 0.90 $\pm$ 0.49 & 1.60 $\pm$ 0.28  \\ \hline
Reference (conv.)&  1.35 $\pm$ 0.28 & 2.03 $\pm$ 1.20 & 3.19 $\pm$ 0.61  \\ \hline
High Current (cond.)  & 0.85 $\pm$ 0.19 & / & 1.17 $\pm$ 0.59 \\ \hline
High Current (conv.) & 1.72 $\pm$ 0.42 & / & 2.69 $\pm$ 1.45 \\ \hline
Low field (cond.) & / & 0.95 $\pm$ 0.17 & 1.50 $\pm$ 0.55 \\ \hline
Low field (conv.) & / & 1.45 $\pm$ 0.29 & 2.86 $\pm$ 1.15 \\ \hline
\end{tabular}
\label{tab2}
\end{table}  

The upstream near SOL heat flux profile appears to have a modest dependence on the plasma conditions, with $\lambda_{q}$ around 1.5-3cm for the convective model and 0.5-1.5cm for the conductive model. This is compatible with the observation that the dominant dependence of $\lambda_q$ is on the magnetic configuration, and in particular on the poloidal magnetic field \cite{Scarabosio2013,Eich2011,Eich2013}. However, also the line averaged density, or possibly the collisionality, seem to have a moderate but visible effect on $\lambda_q$, which somewhat increases with $\overline{n}_e$ or $\nu_{*,e}$.   

\section{Target measurements}

The radial profiles of both the target ion saturation current and parallel heat flux were measured directly with Langmuir probes and infra red thermography respectively. All the outer divertor data were analysed using the same procedure, based on a fitting function that assumes a diffusive broadening of exponentially decaying upstream profiles \cite{Wagner1985,Eich2011}. Its form is given by:
\begin{equation}
\label{1} f(x) = \frac{f_0}{2}e^{\left(\frac{S_f}{2 \lambda_f F_{exp}}\right)^2-\frac{x-x_0}{\lambda_f F_{exp}}}erfc\left(\frac{S_f}{2\lambda_f F_{exp}}-\frac{x-x_0}{S_f}\right)+f_{bg}, 
\end{equation} 
where $f$ can represent the parallel heat flux or the ion saturation current, $S_f$ is the divertor spreading factor and $F_{exp}$ is the target flux expansion. The fitting parameters $f_0$, $f_{bg}$ and $x_0$ measure the amplitude of the field, its background value and the position of the strike point. Note that, in this model, $\lambda_f$ represents the projected \textit{upstream} decay length, which we use as a reference throughout the paper and is directly comparable with the data presented in Sec.\ref{upstream}. To assure compatibility between upstream and downstream measurements, we have restricted the analysis of the target data to the same time window defined in Sec.\ref{upstream}, even though the divertor diagnostics were active throughout most of the discharge time.  

\begin{figure}
\includegraphics[height=12cm,width=12cm, angle=0]{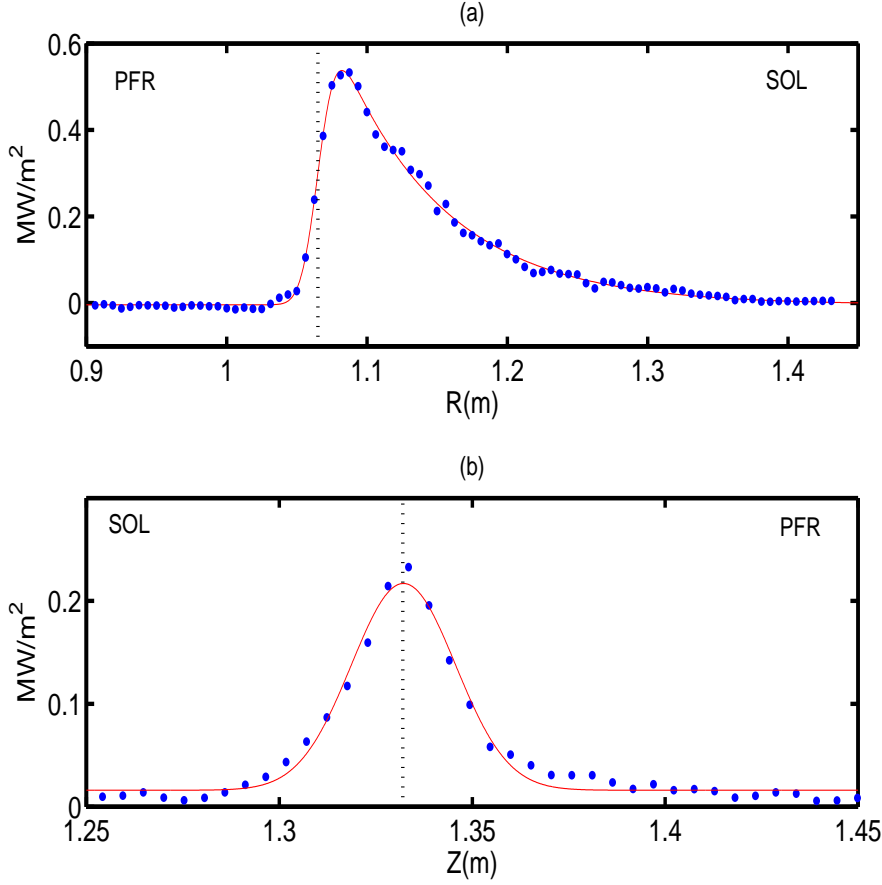}
\caption{Heat flux profiles at the upper outer (a) and inner (b) targets for discharge 28996 at $t=0.2508$ sec. The markers are the IR measurements, while the curves are fits in the form of Eq.\ref{1} for (a) and of Eq.\ref{2} for (b). $R$ is the major radius at the divertor plate and $Z$ is the vertical coordinate on the central solenoid. Dotted lines represent the position of the strike point.}
\label{fig8}
\end{figure}

Figure \ref{fig8}(a) shows a typical profile of the parallel heat flux (averaged over transient events caused by filaments) and its fitting curve which, as also observed in other machines \cite{Eich2013,Makowski2012}, usually provides a very good match. This is not the case, however, for the profiles in the inner target, which require a different approach, as discussed in Sec.\ref{inner_target}. 

\subsection{Ion saturation current}

To increase accuracy, the analysis of the $J_{sat,t}$ profiles was restricted to measurements with a relative error below 15\%. The natural sweep of the strike point was used to construct a better resolved profile by overlapping the profiles at different time frames, after correcting for their increasing radial shift. This was done during the flat top phase and in a 30-50 msec window. The upper outer divertor decay lengths and divertor spreading parameters obtained with this method are summarized in Tables \ref{tab3} and \ref{tab4}.  
\begin{table}
\caption{Ion saturation current decay lengths, $\lambda_J$, calculated at the upper outer divertor (in cm). Fitting errors calculated using a 95\% confidence value.}
\begin{tabular}{|l||c|c|c|}
\hline
& low $\overline{n}_e$ & intermediate  $\overline{n}_e$ & high $\overline{n}_e$ \\
\hline\hline
Reference &  1.37 $\pm$ 0.03 & 2.71 $\pm$ 0.03 & 2.77 $\pm$ 0.05  \\ \hline
High Current  & 1.54 $\pm$ 0.04 & 1.79 $\pm$ 0.03 & 2.57 $\pm$ 0.06 \\ \hline
Low field  & / & 1.46 $\pm$ 0.02 & 2.62 $\pm$ 0.02 \\ \hline
\end{tabular}
\label{tab3}
\end{table}  
\begin{table}
\caption{Same as Tab.\ref{tab3} for the ion saturation current divertor spreading parameter, $S_J$ (in cm).}
\begin{tabular}{|l||c|c|c|}
\hline
& low $\overline{n}_e$ & intermediate  $\overline{n}_e$ & high $\overline{n}_e$ \\
\hline\hline
Reference &  0.74 $\pm$ 0.02 & 0.53 $\pm$ 0.01 & 0.62 $\pm$ 0.02  \\ \hline
High Current  & 0.88 $\pm$ 0.03 & 0.52 $\pm$ 0.02 & 0.61 $\pm$ 0.02 \\ \hline
Low field  & / & 0.66 $\pm$ 0.01 & 0.56 $\pm$ 0.01 \\ \hline
\end{tabular}
\label{tab4}
\end{table}  

\subsection{Parallel heat flux} \label{outer_target}

The good temporal and spatial resolution of the IR system allowed us to treat these data in a different way with respect to the ion saturation current. In particular, we fitted each time frame, thus obtaining the detailed time evolution of both $\lambda_q$ and $S_q$. 

It is important to remark that in all our discharges we observed small sawteeth oscillations. This has interesting consequences for the deposition pattern at the divertor targets, producing a dynamic behaviour in several physical quantities, including $\lambda_q$. A detailed discussion of these effects will be presented in a companion paper \cite{Militello2015}. Despite the transient behaviour induced by the sawteeth, it was possible to extract measurements representative of steady state condition. Indeed, while the upstream density and especially temperature show an oscillating behaviour, the deposition profiles remained self-similar during the sawtooth cycle, apart from a relatively short period of time during the crash ($\sim 1 msec$). This suggests that both $\lambda_q$ and $S_q$ have a low sensitivity to the thermodynamic properties of the upstream plasma.

This is confirmed by comparing the decay lengths of the different discharges performed, all of which show a similar value of $\lambda_q$. For comparison, the average $\lambda_q$ and $S_q$, obtained by removing the data corresponding to the crash phase of the sawtooth, are summarized in Tables \ref{tab5} and \ref{tab6}.
\begin{table}
\caption{Parallel heat flux decay lengths, $\lambda_q$, calculated at the upper outer divertor (in cm). Fitting errors calculated using a 95\% confidence value.}
\begin{tabular}{|l||c|c|c|}
\hline
& low $\overline{n}_e$ & intermediate  $\overline{n}_e$ & high $\overline{n}_e$ \\
\hline\hline
Reference &  1.83 $\pm$ 0.12 & 1.90 $\pm$ 0.17 & 1.96 $\pm$ 0.22  \\ \hline
High Current  & 1.88 $\pm$ 0.16 & 1.97 $\pm$ 0.08 & 2.13 $\pm$ 0.15 \\ \hline
Low field  & / & 2.10 $\pm$ 0.12 & 2.02 $\pm$ 0.20 \\ \hline
\end{tabular}
\label{tab5}
\end{table}  
\begin{table}
\caption{Same as Tab.\ref{tab5} for the heat flux divertor spreading parameter, $S_q$ (in cm).}
\begin{tabular}{|l||c|c|c|}
\hline
& low $\overline{n}_e$ & intermediate  $\overline{n}_e$ & high $\overline{n}_e$ \\
\hline\hline
Reference &  0.32 $\pm$ 0.04 & 0.81 $\pm$ 0.08 & 0.76 $\pm$ 0.11  \\ \hline
High Current  & 0.36 $\pm$ 0.06 & 0.35 $\pm$ 0.03 & 0.70 $\pm$ 0.07 \\ \hline
Low field  & / & 0.26 $\pm$ 0.03 & 0.92 $\pm$ 0.10 \\ \hline
\end{tabular}
\label{tab6}
\end{table}  

\subsection{Inner divertor measurements} \label{inner_target}

In the double null configurations that we have analysed, only roughly 10\% of the power crosses the separatrix towards the high field side. In the low power environment of the inner target, the IR profiles show a symmetric configuration with respect to the strike point, see Fig.\ref{fig8}(b). A consequence of this is that when the Wagner/Eich function, Eq.\ref{1}, is fitted to the inner target IR data the decay length is poorly constrained (it would lead to large errorbars). This suggests that, differently from the outer target, $\lambda_q$ is smaller than $S_q$, and hence the profile is mostly determined by the mechanisms below the X-point. In our analysis, we fitted the profiles with a simple Gaussian function of the form:
\begin{equation}
\label{2}
f(x) = f_0e^{-\left(\frac{x-x_0}{S_f}\right)^2}+f_{bg},
\end{equation}  
which provided a better fit to the data. Table \ref{tab7} summarizes the measurements taken.
\begin{table}
\caption{Parallel heat flux divertor spreading parameters calculated at the upper inner divertor (in cm). Fitting errors calculated using a 95\% confidence value.}
\begin{tabular}{|l||c|c|c|}
\hline
& low $\overline{n}_e$ & intermediate  $\overline{n}_e$ & high $\overline{n}_e$ \\
\hline\hline
Reference &  0.98 $\pm$ 0.09 & 0.79 $\pm$ 0.08 & 1.28 $\pm$ 0.20  \\ \hline
High Current  & 1.03 $\pm$ 0.08 & 0.99 $\pm$ 0.06 & 1.02 $\pm$ 0.05 \\ \hline
Low field  & / & 0.85 $\pm$ 0.06 & 0.76 $\pm$ 0.12 \\ \hline
\end{tabular}
\label{tab7}
\end{table}  

\section{Comparison between upstream and target measurements}

\begin{figure}
\includegraphics[height=12cm,width=12cm, angle=0]{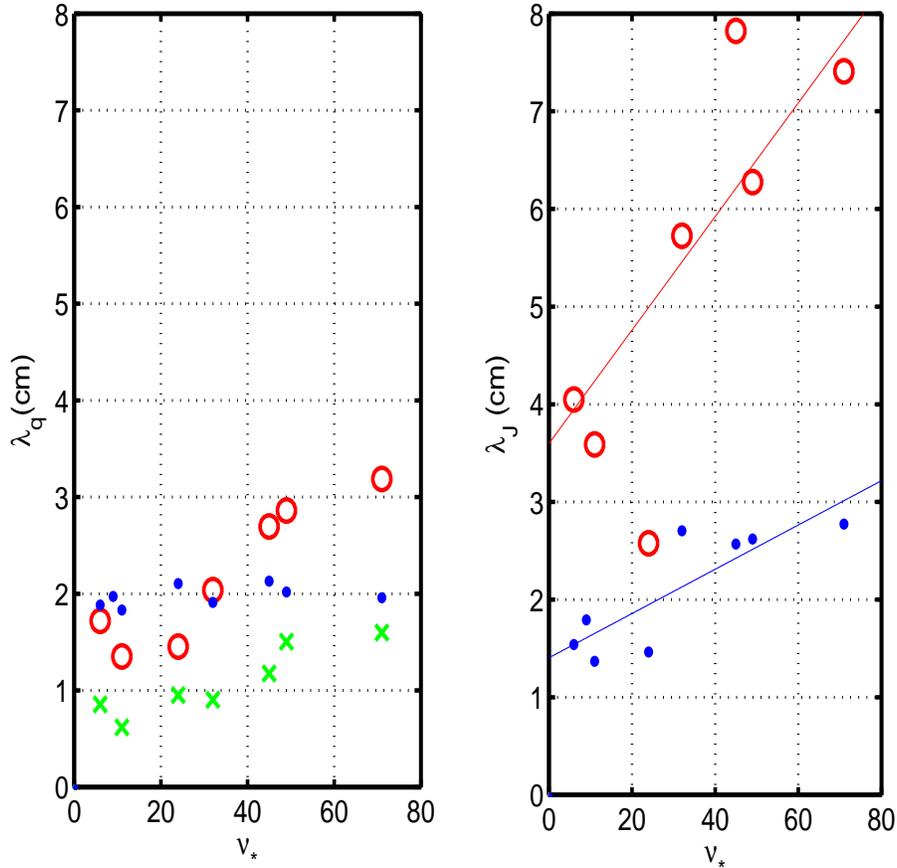}
\caption{Parallel heat flux (left panel) and ion saturation current (right panel) near SOL decay lengths as a function of the electron collisionality. The symbol ($\color{blue} \bullet$) represents the upstream projection of the target measurements, ($\color{red} \circ$) the upstream measurements (with convective assumption for the heat flux) and ($\color{green} \times$) the upstream measurements with conductive assumption for the heat flux. The solid curves on the right panel show the linear fit of the upstream and target projected data.}
\label{fig10}
\end{figure}

Comparing upstream and downstream measurements allows us to gain some insight into the mechanisms that transfer particles and energy towards the divertor. In particular, the downstream data are now commonly interpreted with analysis techniques that assume cross field diffusion \cite{Wagner1985} from a midplane profile (interestingly, they also postulate a \textit{convective} parallel transport for the energy). Figure \ref{fig10} shows that midplane and target measurements are globally in good agreement for the heat flux decay length, while they are not for $\lambda_J$. 

These results question the validity of the standard diffusive approach for the particle cross field transport in the SOL, which is likely to be governed by the complex motion of high density filamentary structures that are observed in all our discharges. It is also interesting to notice that the near SOL $\lambda_J$ has a well defined increasing trend with the upstream collisionality, both at the midplane and at the target. 

In Fig.\ref{fig10}, the heat flux decay lengths upstream are evaluated with convective and conductive assumptions, with the former displaying a better agreement with the projected target data for low collisionality and vice versa for the latter. Overall, both upstream estimates show an increase of $\lambda_{q,u}$ with the collisionality. On the other hand, the projected $\lambda_q$ is remarkably constant in different experimental conditions, which suggests that this quantity has a weak dependence on both thermodynamic and magnetic properties of the plasma. In addition, it seems like the upstream density broadening does not affect significantly the heat flux target profiles. A proper estimation of the scaling factors would require a database extended to more than 8 discharges (e.g. see \cite{Eich2013}).      

\begin{figure}
\includegraphics[height=12cm,width=12cm, angle=0]{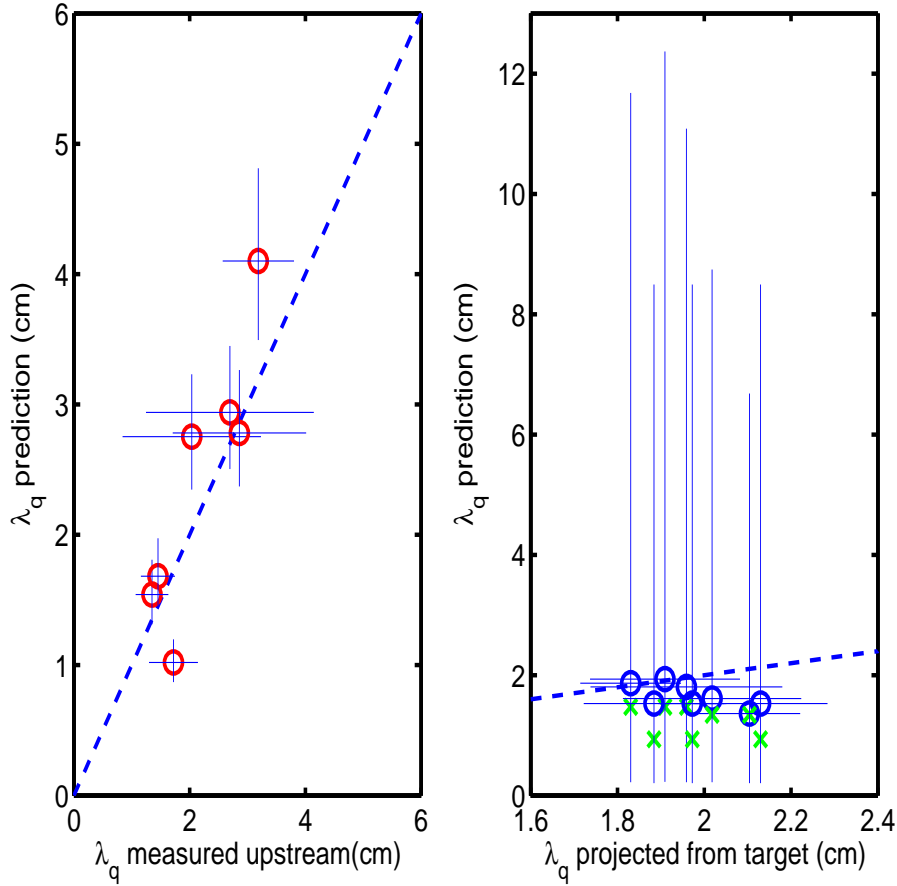}
\caption{Left panel: comparison between the parallel heat flux measured upstream and the scaling law of \cite{Militello2013} ($\color{red} \circ$, theoretical). Right panel: comparison between the upstream parallel heat flux projected from the target measurements and the scaling laws of \cite{Scarabosio2013} ($\color{blue} \circ$, experimental) and of \cite{Goldston2012} ($\color{green} \times$, theoretical). the solid lines are the errorbars on the measurements or on the predictions. The thin dashed lines represent the diagonals.}
\label{fig11}
\end{figure}

In this respect, it is useful to compare our measurements with recent experimental or theoretical scaling laws of $\lambda_q$. The multi-machine regression based on infra red analysis performed in \cite{Scarabosio2013} predicts a decay length of 1.4-1.9cm for our dataset, but with large error bars that can go one order of magnitude above or below the nominal value. The theoretical scaling of \cite{Militello2013}, obtained with midplane 2D turbulence simulations, provides a good fit to the the convective upstream measurements, especially considering the low errorbars on the theoretical predictions (corresponding to a relative error of 15\%). An exception is the low density high current case, which is characterised by the lowest collisionality of the dataset. This might suggest that the flux limiter closure used in \cite{Militello2013} is not sufficient to capture the parallel dynamics and non-local approaches \cite{Omotani2013} might be needed to improve the agreement. However, the matching of this midplane model is not as satisfactory where the projected target data are concerned. An heuristic theoretical model based on drifts was proposed in \cite{Goldston2012} and its predictions give a $\lambda_q$ around 0.9-1.45cm. An important caveat is that the Goldston model \cite{Goldston2012} was developed for H-mode plasmas. A summary of these comparisons is shown in Fig.\ref{fig11}.      

\section{Summary and conclusions}

In this paper, we discussed a series of experiments carried out in the MAST spherical tokamak aimed at characterising the behaviour of the SOL decay lengths as a function of the plasma conditions. Our results show that the density and parallel heat flux profiles respond to changes in the plasma in a different way, both at the midplane and at the target.

Only the particle exhaust, for example, is significantly affected by the density (or the fuelling), a higher level of which corresponds to broader density profiles in the SOL. Such broadening occurs both in the \textit{near} SOL, where the exponential decay becomes less steep, and in the \textit{far} SOL where, in certain cases, the formation of a shoulder in the density profile is observed. The experiments that we performed showed that the far SOL flattening can occur even in the absence of divertor detachment or wall recycling, although both mechanisms are likely to play a role in particular cases. It is interesting that even 2D turbulence models without divertor or neutral particle physics capture the profile broadening \cite{Militello2012,Militello2013}, although they also suggest that a proper 3D treatment in realistic geometry is required to unravel the details behind it. While the 3D dynamics of the filaments were addressed by the community only recently \cite{Angus2012,Angus2012b,Halpern2014,Easy2014,Omotani2015}, the first results already anticipate a rich phenomenology in which both local and global features of the plasma filaments can affect their motion in the SOL. In particular, crucial effects could include the presence and structure of the X-point; the parallel profile of the temperature (or collisonality) in the filament; the electrical connection of the filament to the Debye sheath at the target through the background plasma or its absence; the parallel and perpendicular shape and dimension of the filament; the ionization in the filament caused by its higher temperature with respect to the background. While the divertor regime (its collisonality, the occurrence of detachment) certainly affects some of the items above, it does not provide a complete picture. Similarly, wall recycling changes the particle source, which might be important for machines with a narrow gap between separatrix and wall, but does not have the generality to capture the phenomena observed in MAST.      

It is interesting to observe that the shoulder formation is not observed in our high current and low density low magnetic field discharges, which suggests that the safety factor could play a role in this phenomenon. On the other hand, we do not rule out the possibility that the broadening can occur at larger distances from the separatrix in these latter cases (i.e. the far SOL is farther away from the separatrix). In our experiments we did not monitor the SOL at distances beyond $\sim 5cm$ as this would imply analysing reciprocating probe data corresponding to a different plasma regime (since the probe plunges at a certain velocity in the plasma, the far SOL corresponds to earlier times in the discharge). From a practical point of view, however, this makes little difference, since the density profile would anyway reach low levels before flattening, thus limiting the plasma wall interactions.   

From an experimental point of view, pinpointing what exactly causes the flattening on the near and far SOL proved extremely difficult. This was due to the limited database and the uncertainty in the measurements, notably the significant errorbars on the position of the separatrix ($\sim$ 1cm) and on the edge temperature (relative error $\leq 65\%$). Both these quantities affect the determination of the collisonality $\nu_*\sim L_\parallel/T^2$, which is a scaling parameter often used to interpret and extrapolate the SOL particle exhaust. While more precise measurements of the upstream temperature are desirable and to a certain extent already possible using other diagnostics, the quadratic dependence of the collisonality on this quantity makes it problematic anyway. In addition, another degree of arbitrariness is introduced by the singular behaviour of the safety factor on the last closed flux surface, which allows a continuum of connection lengths (and therefore collisionalities) depending on the radial distance from the separatrix where this quantity is estimated. In order to resolve this conundrum, only the combination of better measurements with first principle understanding of the SOL mechanisms is a viable solution.

Differently from the particle exhaust, the energy exhaust displays a remarkable insensitivity to the thermodynamic state of the main plasma, at least at the target. This is in agreement with experimental multi-machine scaling laws \cite{Scarabosio2013}, which relate the parallel heat flux decay length mainly to the magnetic structure of the plasma. As a matter of fact, our target data are well represented by the L-mode Scarabosio scaling. The weak dependence of the target decay length on the upstream density and temperature is further confirmed by its response to sawtooth oscillations. We consistently observed a self-similar profile in the divertor IR data during the sawtooth cycle, apart from a short period around the crash when the target profile has to dynamically respond to the higher heat flux received. Our analysis cannot clarify if the absence of thermodynamic effects on the target $\lambda_q$ implies that magnetic geometry dominates over plasma physics or that there is an almost complete cancellation of the latter. The upstream heat flux decay length seems to be more sensitive, albeit only moderately, to the local density and temperature and are reproduced within error bars by theoretical scaling laws based on upstream turbulence \cite{Militello2013}. The inner target heat flux deposition features a Gaussian profile rather than the convolution between a Gaussian and an exponential used in the outer divertor. This suggests a quite different in/out exhaust mechanism, at least in the double null configurations examined.     

The results above suggest that the particle and energy exhaust occur through different, albeit probably related, mechanisms. This is further confirmed by the fact that the midplane projection of the divertor data, based on perpendicular diffusion models, gives a good match with the measured profiles of the heat flux but not of the saturation current (i.e. the density). A possible explanation for this discrepancy relies on the different parallel transport paradigms followed by the particle and energy channels \cite{Fundamenski2007}. The heat is efficiently exhausted along field lines by the electrons, which are highly conductive in the parallel direction. This means that, whatever the mechanism that moves the energy across field lines (turbulent eddies, filaments, collisions), it has little time to act before most of the heat is drained at the divertor. Conversely, the particle exhaust in the parallel direction is advective and much slower, so that the characteristic length scale for the SOL density, in the range of collisionality considered, is larger than that of the heat flux. In conclusion, as the steep heat flux profile samples mainly the diffusive region close to the separatrix, the projection technique works well, while it does not for the particle exhaust, characterised by non-local non-diffusive transport related to the filaments.

Finally, the limited number of discharges performed did not allow us to extract reliable scaling laws from the data collected. This explains the mostly qualitative nature of our discussion and advocates for multi-machine analysis in order to provide quantitative relations between the relevant parameters of the problem.   

\section{Acknowledgements}
F.M. acknowledges useful discussions with Prof. B. Lipschultz, Dr. S. Elmore and Mr. N. Walkden. We aslo thank Dr. K. Lawson for carefully reading the manuscript. This work has been carried out within the framework of the EUROfusion Consortium and has received funding from the Euratom research and training programme 2014-2018 under grant agreement No 633053 and from the RCUK Energy Programme [grant number EP/I501045]. To obtain further information on the data and models underlying this paper please contact PublicationsManager@ccfe.ac.uk. The views and opinions expressed herein do not necessarily reflect those of the European Commission.

\end{document}